# Negative refractive index and higher-order harmonics in layered metallodielectric optical metamaterials

R. Maas, E. Verhagen, J. Parsons, and A. Polman

Center for Nanophotonics, FOM Institute AMOLF

Science Park 104, 1098 XG Amsterdam, the Netherlands

**Abstract**

We study the propagation of light in a three-dimensional double-periodic Ag/TiO$_2$ multilayer metamaterial composed of coupled plasmonic waveguides operating in the visible and UV spectral range. For these frequencies, light propagation in the plane of the waveguides is described by a negative phase velocity, while for the orthogonal direction light propagation is described by a Bloch wave composed of a large number of harmonics. As a result, the material cannot generally be described by a single phase index: decomposing the Bloch wave into different harmonics we show that for the wavelength range of interest the positive index $m$=1 harmonic dominates the propagation of light in the orthogonal direction. These results are corroborated by numerical simulations and optical refraction experiments on a double-periodic Ag/TiO$_2$ multilayer metamaterial prism in the 380-600 nm spectral range, which show that positive refraction associated with right-handed harmonics dominates. Studying the isofrequency contours we find that despite the occurrence of multiple harmonics the double-periodic structure can act as a flat lens: for a slab consisting of an integer number of unit cells all harmonics are degenerate and constructively interfere at the image plane. This work identifies important considerations relevant to the design of many three dimensional periodic metamaterials.

**Introduction**

Negative-index or "left-handed" metamaterials are structures in which light propagates with opposite phase and energy velocities[1], enabling an entirely new class of optical components such as *e.g.* flat lenses[2], potentially allowing an image resolution below the diffraction limit[3,4,5,6]. In the past decade, negative-index metamaterials have been fabricated for a broad range of frequencies[7]. The first demonstration was achieved in the microwave regime[8], and since then, scaling down of the unit cell size enabled by advances in nanofabrication has led to metamaterials with operation frequencies towards the visible spectral range[9].

Most metamaterial design architectures are based on resonant geometries such as e.g. split rings[10], fishnet structures[11], and others[12]. Recently, our group has proposed a three-dimensional, negative-index metamaterial based on a double-periodic metal/dielectric multilayer geometry, operating in the UV spectral range[13]. It is based on a multilayer stack of negative-index waveguides[14,15,16], where TM-polarized light couples to surface plasmon polaritons at the metal/dielectric interfaces. By careful design of the coupling between adjacent waveguides, an angle-independent negative fundamental mode index can be achieved, implying that light propagation in the plane of the waveguides and normal to it is characterized by the same (negative) propagation constant. An experimental realization of this geometry has been used to study negative refraction of energy and to show flat lensing behavior[17].

To further investigate the behavior of this multilayer metamaterial geometry it is important to evaluate the role of other (higher-order) harmonics, aside from the fundamental harmonic. The solution of the wave equation in a periodic medium is a Floquet-Bloch wave, composed of a superposition of multiple plane waves with different propagation constants. Previous work on photonic crystals shows how the periodicity of the geometry leads to an ambiguity in the determination of an effective phase index[18,19]. In this paper, we study these harmonics and quantify their propagation characteristics, specifically for a metal-dielectric multilayer structure which exhibits strong dispersion and significant absorption in the UV-visible spectral range. First, we use a transfer matrix formalism to determine the exact properties of the Bloch wave in a theoretical infinitely thick periodic medium. The decomposition into harmonics shows that the fundamental harmonic (located in the first Brillouin zone) is not the most dominant in the system. Comparing this decomposition with isofrequency contours (IFCs) we show that the dominant harmonic is not characterized by a left-handed response. Secondly we study the implications of this result for phase and energy refraction in finite structures, with analytical calculations, numerical simulations and finally by performing refraction experiments using a metamaterial prism.

**Field decomposition**

A sketch of the geometry studied here is shown in Figure 1a. A multilayer stack is formed by alternating Ag and $TiO_2$ layers. Light incident on the side of this stack will couple to waveguide modes. The layer thicknesses are optimized such that at frequencies above the surface plasmon resonance frequency there is only a single negative index waveguide mode supported by the system[16]. As light propagates parallel to the waveguides, it experiences a constant environment, and propagates with a negative waveguide mode wavevector. This causes phase and energy to travel in anti-parallel directions, a key characteristic of left-handed materials. The fact that the metal layers are alternatingly thin and thick causes the coupling constant between adjacent unit cells to be negative[13], such that phase and energy velocities are antiparallel for a wide range of angles.

However, for wave propagation in any direction other than parallel to the waveguides, the periodically varying permittivity experienced by light physically changes the nature of wave propagation. To study this, we consider the extreme limit of light travelling normal to the waveguides. Using a transfer matrix method, we calculate how light propagates through an infinitely periodic multilayer structure. We adopt the notation of Russell[20], and calculate the field in this periodic unit cell from the transfer matrix, assuming TM polarization. We express the magnetic field profile in a single layer as[21]:

$$H_{y,j}^{N}(z) = a_j^N \cos\left(k_{z,j}\left(z-z_j^N\right)\right) + b_j^N \frac{n_j^2}{ak_{z,j}} \sin\left(k_{z,j}\left(z-z_j^N\right)\right), \quad (1)$$

where $k_{z,j}$ is the wavevector component along $z$, defined by $k_{z,j} = \sqrt{n_j^2 k_0^2 - k_x^2}$, $z_j^N$ is the center of layer $j$ in unit cell $N$, $n_j$ is the refractive index of layer $j$ and $a$ is the unit cell size. From this magnetic field profile also the electric field profile can be determined using the Ampère-Maxwell law. The coefficients $a_j^N$ and $b_j^N$ are related by:

$$\begin{pmatrix} a_j^{N+1} \\ b_j^{N+1} \end{pmatrix} = M_{trans} \begin{pmatrix} a_j^N \\ b_j^N \end{pmatrix}. \quad (2)$$

The coefficients of the first layer in unit cell $N$ can be defined as[20]: $a_1^N = \sqrt{(M_{trans})_{12}}$ and $b_1^N = \pm\sqrt{(M_{trans})_{21}}$, where $M_{trans} = M_{14}M_{43}M_{32}M_{21}$, and the matrix $M_{ji}$ relates the coefficients in adjacent layers through $\begin{pmatrix} a_j^N \\ b_j^N \end{pmatrix} = M_{ji} \begin{pmatrix} a_i^N \\ b_i^N \end{pmatrix}$, such that the boundary conditions are fulfilled. This definition of $M_{trans}$ imposes periodic boundary conditions, leading to an infinitely periodic, closed system. The eigenvalues $\lambda_\pm$ of the transfer matrix are related to the complex Bloch wavevector in the first Brillouin zone as $\lambda_\pm = \exp(\pm ik_{Bl}a)$. In the above, ± refers to a wave with energy travelling in the +z or –z direction. We perform calculations for a Ag/TiO$_2$/Ag/TiO$_2$ double-periodic geometry, using optical constants and layer thicknesses from Xu et al.[17] (see Table 1). We consider waves with energy propagating in the +z direction, such that $\text{Im}(k_{Bl}) \geq 0$. At a wavelength $\lambda_0$=363.8 nm and for normal incidence we find $k_{Bl} = -17.17 + 4.05i \; \mu m^{-1}$, where the negative real part reflects that phase propagates backwards as energy propagates forwards for the fundamental harmonic.

We repeat this calculation for a whole range of wavelengths to determine the band diagram for this geometry. The first Brillouin zone is shown in Fig. 1b. In this calculation optical constants are used determined by variable-angle spectroscopic ellipsometry on individual Ag and $TiO_2$ layers, deposited on a Si substrate using physical vapor deposition (solid lines). From the band diagram we can identify two regimes where the imaginary part of the Bloch wavevector (shown in Fig. 1c) is relatively low, corresponding to weak attenuation. Also shown is the band structure for the same geometry, but with the imaginary part of the permittivity set to zero (dashed lines). Two propagation bands are clearly observed at $\lambda_0$=345-365 nm and $\lambda_0$=400-425 nm, and correspond to the two regimes of low attenuation in the geometry with absorption.

The field profile of the Bloch wave, known from Eq. 1, can be expanded as:

$$H_y(z) = \exp(ik_{Bl}z) \sum_{m=-\infty}^{\infty} h_{y,m} \exp\left(i\frac{2\pi}{a}mz\right) \quad (2)$$

Where $h_{y,m}$ represents the amplitude of the plane wave (harmonic) indicated by index $m$, with wavevector $k_m = k_{Bl} + \frac{2\pi}{a}m$. The amplitude $h_{y,m}$ can be found by performing a simple Fourier decomposition:

$$h_{y,m} = \frac{1}{a}\int_0^a H_y(z)\exp(-ik_{Bl}z)\exp\left(-i\frac{2\pi}{a}mz\right)dz \quad (2)$$

This shows that only considering the first Brillouin zone, as in Fig. 1b, does not capture the full propagation characteristics of the Bloch wave. Instead, all harmonics with their corresponding weights should be taken into account. The different harmonics can be plotted in a wavevector diagram, where the Bloch wavevector is calculated at a fixed frequency, but now for different angles of propagation (so different values of $k_x$). The interface boundary conditions are angle dependent, leading to a change in the Bloch wavevector. In the calculation, the parallel wavevector component $k_x$ is chosen to be real valued.

The wavevector diagram is given in Fig. 1d, where we show the real part of $k_z$ plotted against $k_x$, at a free-space wavelength of 364 nm. Here we include not only the fundamental isofrequency contour, but also that of the first higher orders (m=-1,0,1; red, green, blue). From the band diagram (Fig. 1b) we determine that the time and unit cell averaged Poynting vector is directed inwards[13,22] for this frequency. The bold isofrequency contours correspond to the Bloch wave propagating in the positive z direction. The *m*=0 and *m*=-1,-2,… IFCs in Fig. 1d show left-handed behavior as wavevector $k_m$ and Poynting vector *S* (indicated by arrows in the figure) are antiparallel. In contrast, the *m*=1,2,… IFCs do not show such behavior and do not satisfy the criterion $k_m \cdot S < 0$. Thus, $m \leq 0$ harmonics can be called "left handed", and *m*>0 "right handed".

From the wavevector diagram we have identified that different harmonics can be associated with either a left- or right-handed response. Therefore it is important to know the exact contribution of each harmonic to the total Bloch wave. We determine these contributions by decomposing the Bloch wave using Eq. 4. The result of this decomposition is shown in Figure 2 for the range -10<*m*<10. As is clear from this figure, a large number of harmonics have a significant amplitude. This implies that to properly describe light propagation in this multilayer metamaterial geometry a large number of harmonics must be taken into account. This is contrary to earlier work where only the fundamental *m*=0 harmonic was considered[23,24]. Interestingly, the *m*=1 harmonic has an amplitude higher than that of the fundamental harmonic. We performed the same analysis for the electric field component $E_x$; the amplitude coefficients for these harmonics are also shown in Fig. 2, a similar trend is found.

The fact that a large number of harmonics contribute significantly to the Bloch wave implies that the phase index of the structure cannot be unambiguously defined: each harmonic has a different propagation wavevector $k_m$, and therefore a different phase velocity $v_{ph,m} = \frac{\omega}{k_m}$. Note that the Bloch wavevector is always complex valued (we write $k_m = k_m' + k_{Bl}''$), such that the phase velocity is also complex valued. As the propagation wavevector always appears in an exponential term, the imaginary part corresponds to an attenuation factor $\exp(-k_{Bl}'' z)$, and the real part determines the phasor term $\exp(ik_m' z)$. Therefore, the phase velocity can be defined as $v_{ph,m} = \dfrac{\omega}{k_m'}$.

To determine the energy density carried by the different harmonics, we first calculate the energy velocity[25] $v_e = \dfrac{\langle\langle S(z)\rangle_t\rangle_{UC}}{\langle\langle U(z)\rangle_t\rangle_{UC}}$. We determine the time and unit cell averaged Poynting vector S and energy density U of the Bloch wave from the $E_x(z)$ and $H_y(z)$ field profiles. Because there is significant dispersion and loss, we use $U(z) = \dfrac{1}{4}\left(\dfrac{\partial(\omega\varepsilon(\omega,z)\varepsilon_0)}{\partial\omega}\left|\vec{E}(z)\right|^2 + \mu\mu_0\left|\vec{H}(z)\right|^2\right)$, from Ref. (26)[26]. The result is shown in Fig. 3: the energy velocity is always positive, and increases in the regions of low loss. The calculated Poynting vector is a property of the complete Bloch wave, and not of individual harmonics. This is due to the fact that the calculation of the Poynting vector contains a multiplication of *E* and *H* fields, leading to cross terms relating different harmonics. However, integrating the Poynting vector over the unit cell leads to a cancellation of these cross terms, so the unit-cell averaged Poynting vector can in fact be decomposed into separate harmonics. Performing this decomposition we find that the *m=1* harmonic component has the highest amplitude, similar to what we found for the decomposition of the $E_x$ and $H_y$ field profiles. The result of the unit-cell averaged Poynting vector decomposition is also included in Fig. 2. The absolute value of the components is divided by the energy velocity to arrive at the energy density.

The *m*=1 component carries 44% of the energy, whereas the fundamental *m*=0 component only accounts for 0.1% of the total energy carried by the Bloch wave. All right-handed components of the Bloch wave combined contain 62% of the wave energy, see Table 2. The fact that multiple higher-order harmonics carry a significant amount of the energy indicates that the structure cannot be described by a single refractive index. Figure 4 shows the total energy fraction carried by all right-handed harmonics combined as a function of wavelength. As can be seen, this fraction is larger than 50% for almost all wavelengths, only in the regime where attenuation is very strong the left-handed harmonics carry most of the energy.

**Finite sized multilayer structures**

In the above, we have analytically studied light propagation through an infinitely periodic multilayer structure. Next, we shall investigate how light interacts with different finite sized structures, to show that the above considerations have a pronounced impact on the optical response of finite sized multilayer structures. First, we perform exact transfer-matrix calculations of the phase accrue of a normal-incident plane wave transmitted through a Ag/TiO$_2$ multilayer slab. This phase advance is determined at the vacuum side of the exit interface, relative to the phase at the vacuum side of the first interface. The transmitted phase is calculated for a slab with a gradually increasing thickness. Figure 5a shows calculations for three different wavelengths. For a wavelength of $\lambda_0$=364 nm and $\lambda_0$=410 nm the phase advance in Fig. 5a has a positive overall slope. At a wavelength of $\lambda_0$=450 nm there is no overall increase of the transmitted phase.

As we have seen in the periodic structure, a large number of harmonics contribute to wave propagation. Here, the phase increases along a single slope, with small deviations due to multiple internal reflections in individual layers. If we approximate this phase advance by a single effective wavevector, we can fit a linear function $\varphi(L) = k_{eff} L$ to the data of Fig. 5a and similar calculations in the 0.3-0.5 $a/\lambda_0$ frequency range. The results are shown as solid dots in the band diagram of the metamaterial (Fig. 5b). The colors of the bands correspond to the different harmonics ($m$=-1,0,1). As can be seen for longer wavelength ($a/\lambda_0$<0.4) the $m$=0 harmonic is dominant, however, at these wavelength the fundamental harmonic is described by right-handed behavior. For shorter wavelengths ($a/\lambda_0$>0.4) The fundamental harmonic is characterized by a left-handed response, but the (right-handed) $m$=1 harmonic dominates. This is in good agreement with the decomposition of the Bloch wave for different wavelengths; the overall phase advance is determined by the harmonic with the highest amplitude.

At normal incidence, the harmonic wavevectors are all aligned. However, when the Bloch wave meets an interface at an angle[27], refraction is determined by the conservation of the parallel projection of the wavevector on this interface (Fig. 6a). Therefore each harmonic will refract with a different angle. Using finite-difference time-domain simulations we study a Ag/TiO$_2$ prism composed of 15 unit cells, truncated by an output facet at an angle of 5 degrees; layer thicknesses and optical constants are as in Table 1. Figure 6b shows the far-field angular distribution of the electric field intensity. For a plane wave at normal incidence multiple refraction peaks are observed, corresponding to the different harmonics. The expected refraction angles from the different harmonics are constructed using the IFCs in Fig. 1d, and are indicated by vertical dashed lines. The simulated refraction peaks correspond very well to these constructed refraction angles. We note that the refraction angles are the same as the diffraction orders from the grating formed by our output facet. While the grating could thus in principle redistribute the various harmonics in the metamaterial among the different refraction peaks, it is striking that the peak amplitude distribution is similar to the harmonic decomposition of the electric field shown in Fig. 2. The fact that the *m*=1 harmonic is the dominant refraction component from the metamaterial prism is a strong indication for the dominance of this harmonic in the internal Bloch wave.

In order to study refraction experimentally, a 6 μm × 6 μm microprism was sculpted from a Ag/TiO$_2$ multilayer stack. This multilayer stack consisted of 2.5 periods of a Ag/TiO$_2$/Ag/TiO$_2$ unit cell with layer thicknesses given in Table 1, optimized for operation at a free-space wavelength of 410 nm. This stack was deposited on a freestanding SiN membrane using physical vapor deposition. Using focused ion beam milling with a 30 kV Ga$^+$ beam at an oblique angle, an output facet with an angle of 4.1° with respect to the layers was fabricated, forming a prism in the multilayer stack. A scanning electron micrograph side view of this output facet is shown in the inset of Figure 7b.

Refraction by the metamaterial sample was measured in a Fourier microscope (Fig. 7a). Spectrally filtered light from a Xe-arc lamp was incident on the flat side of the metamaterial prism. The transmitted light was collected using a 20x 0.45 N.A. microscope objective, and first spatially filtered using a pinhole in the image plane of the collection objective. The back-focal plane of the objective was then directly imaged onto a Si CCD detector. Figure 7b shows the measured normalized angular distribution of refracted light relative to the normal of the output facet. The angular distribution is converted to an effective prism phase index inside the metamaterial using Snell's law. As can be seen, most light is refracted to angles that correspond to a positive effective phase index (ranging from 1.5>$n_{eff}$>0.5) over the 370-570 nm spectral range: for decreasing wavelength the index is increasing to progressively larger values in a stepwise manner similar to the calculation in Fig. 5b.

Finally, we note that the fact that the phase index cannot be uniquely defined for the multilayer stack does not prevent flat lensing[27,28,29,30]. The spatial phase profile at the exit interface of the metamaterial that results in the construction of an image at a position beyond the slab of a point source at the entrance interface is the same for every harmonic, as it is related only to the curvature of the IFC, which is the same for every harmonic; see Fig. 1d. This is comparable to using the curvature of the hyperbolic dispersion of anisotropic right-handed media for imaging[27,30]. Flat lensing, which has been experimentally observed by Xu *et al.* for the layer geometry studied here, thus is the result of image reconstruction due to the coherent superposition of refracted light from multiple harmonics in the metamaterial. The image is formed due to negative refraction of energy, invoked by the shape of the isofrequency contour, and not by the left-handed nature of the fundamental harmonic. As demonstrated above, light propagation is dominated by right-handed harmonics with a positive phase index.

**Conclusions**

We have studied the propagation of light in the 300-500 nm spectral range in a double-periodic Ag/TiO$_2$ multilayer metamaterial. Based on calculations and simulations we conclude that light propagation in the metamaterial is governed by multiple harmonics with either a positive or negative phase index. We find that positive index ("right-handed") harmonics dominate in the wavelength range of interest. More than half of the energy is carried by positive-index harmonics. Flat lensing in the material is the result of the coherent superposition of multiple refracted harmonics, dominated by "right-handed" harmonics, and is similar in nature to that of hyperbolic metamaterials. Nonetheless, there is a marked difference between hyperbolic metamaterials and the geometry considered here; the effective optical properties are determined by coupling to negative index waveguide modes. Our work is confirmed by experiments and numerical simulations on a multilayer metamaterial microprism from which the multiple refracting harmonics are directly resolved, with a positive-index harmonic dominating. The striking differences between individual harmonics of the same Bloch waves require careful consideration, not only for the metamaterial design outlined in this work, but more generally for other periodic metamaterials as well.

**Acknowledgements**

The authors would like to acknowledge Femius Koenderink for useful discussions. This work is part of the research programme of the Foundation for Fundamental Research on Matter (FOM), which is financially supported by The Netherlands Organization for Scientific Research (NWO). It is also funded by the European Research Council.

|  | Model | Experiment |
|---|---|---|
| $d_1$ TiO$_2$ | 28 nm | 44 nm |
| $d_2$ Ag | 30 nm | 32 nm |
| $d_3$ TiO$_2$ | 28 nm | 44 nm |
| $d_4$ Ag | 66 nm | 50 nm |
| $\varepsilon_{Ag}$ (364 nm) | -2.522+0.250i | -2.214+0.262i |
| $\varepsilon_{TiO2}$ (364 nm) | 7.838+0.280i | 7.566+0.175i |

**Table 1** Optical constants and layer thicknesses used in the model and in the experiment. Layer thicknesses and optical constants in the model are taken from Ref. (17). The layer thicknesses of the multilayer structure in the experiment where optimized for operation at a free-space wavelength of $\lambda_0$=410 nm. The optical constants of deposited Ag and TiO$_2$ layers were experimentally determined using variable-angle spectroscopic ellipsometry.

| m  | $k_m$ (μm$^{-1}$) | $|S_m|/S_{total}$ |
|----|-------------------|-------------------|
| -2 | -99.85+4.05i      | 0.0086            |
| -1 | -58.51+4.05i      | 0.3612            |
| 0  | -17.17+4.05i      | 0.0009            |
| 1  | 24.16+4.05i       | 0.4386            |
| 2  | 65.50+4.05i       | 0.1661            |

**Table 2** Wave vector $k_m$ and unit cell averaged Poynting vector amplitude decomposition $|S_m|$ of different harmonics of the Bloch wave, calculated for a free-space wavelength of 363.8 nm, using the layer thicknesses and material constants in Table 1. $S_{total}$ is defined as: $S_{total} = \sum_{m=-\infty}^{\infty} |S_m|$. The positive index *m*=1 harmonic has the highest amplitude. Furthermore, the sum of $|S_m|/S_{total}$ for *m*=1,2,… equals 62%, indicating that most of the energy of the Bloch wave is carried by right-handed harmonics.

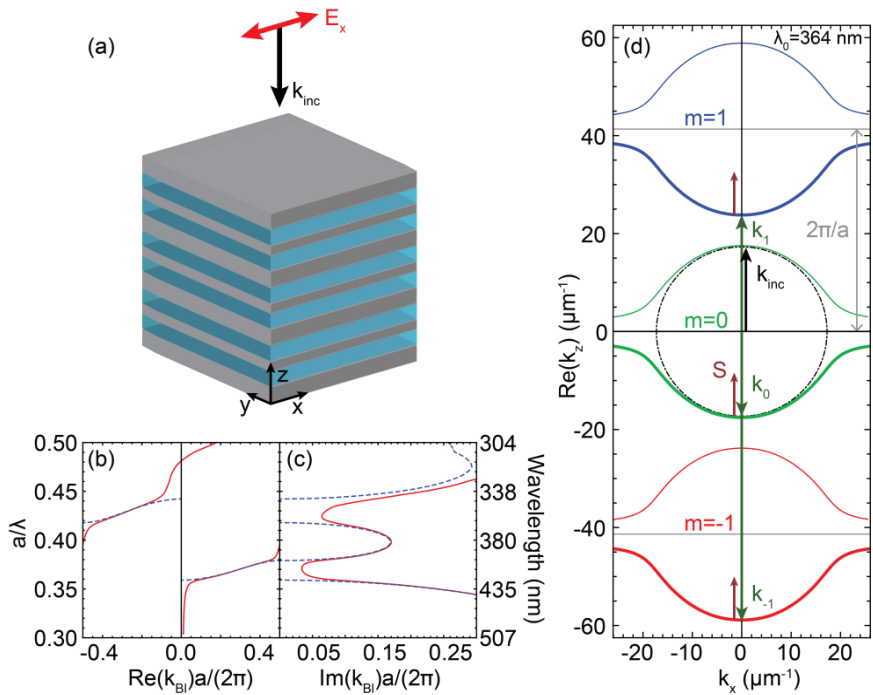

**Figure 1 a**, Sketch of the experimental double periodic geometry. Gray layers correspond to Ag and blue layers to $TiO_2$. **b**, Calculated band diagram of the double-periodic metamaterial. The unit cell size is $a$=152 nm. Solid (dashed) lines correspond to the geometry with (without) losses. Wave frequency (in units of $a/\lambda_0$) is plotted versus the real (b) and imaginary (c) part of the Bloch wavevector. The sign of the wavevector is chosen such that Im($k_{Bl}$)>0 at every frequency, corresponding to the forward propagation direction. Two transmission bands are observed at $\lambda_0$=345-365 nm and $\lambda_0$=400-425 nm. **d**, Isofrequency contours for the Ag/$TiO_2$ double-periodic multilayer metamaterial (*x* and *z* are in-plane and normal components, respectively). Data are shown for fundamental (*m*=0) and higher-order (*m*=-1,1) harmonics at a free space wavelength of $\lambda_0$=364 nm. The Poynting vector is normal to the isofrequency contour with the direction determined by the sign of d$\omega$/dk derived from IFC calculations at multiple frequencies. Bold-plotted IFC branches correspond to energy propagation in the positive *z* direction. A negative refractive index occurs when the wavevector is antiparallel to the time-averaged Poynting vector which is the case for the *m*=-1,0 harmonics but not for the dominant *m*=1 mode.

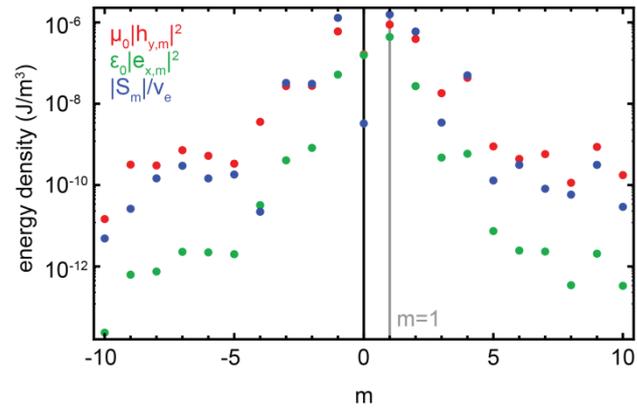

**Figure 2** Calculated amplitude of harmonics in the range -10<*m*<10 for the periodic electric and magnetic field profiles at a free space wavelength of $\lambda_0$=364 nm. The decomposition of the unit cell averaged Poynting vector is also shown. A large number of harmonics have a significant amplitude, and for all decompositions the *m*=1 harmonics has the highest amplitude.

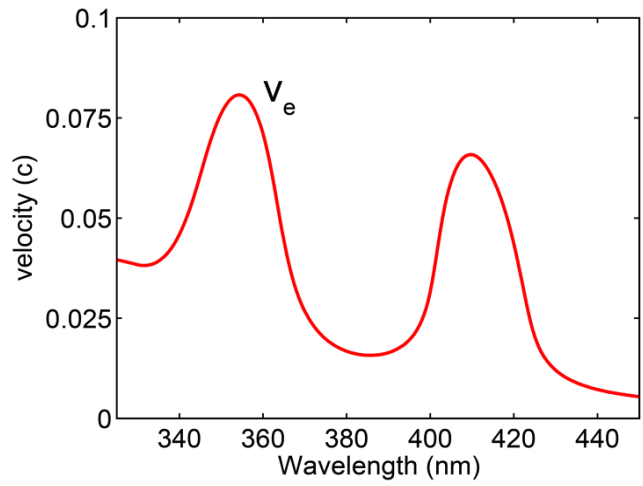

**Figure 3** Calculated energy velocity for the multilayered metamaterial. The energy velocity is significantly higher in the two propagation bands where the attenuation is relatively low: $\lambda_0$=345-365 nm and $\lambda_0$=400-425 nm.

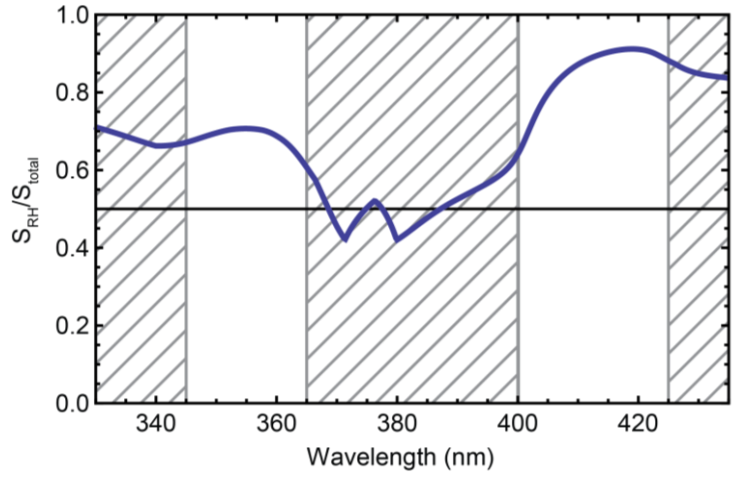

**Figure 4** Calculated total energy fraction carried by all right-handed harmonics combined, as a function of wavelength. Shaded regions correspond to wavelength at which the multilayer structure shows strong absorption.

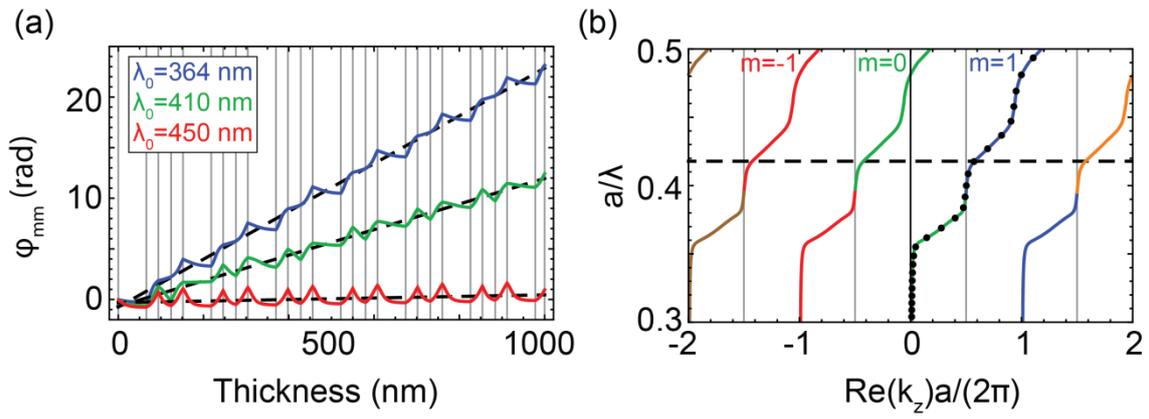

**Figure 5 a**, Calculated phase shift as a function of multilayer stack thickness for free-space wavelengths of $\lambda_0$=364, 410 and 450 nm. The vertical gray lines indicate the metal/dielectric interfaces. A linear function (dashed black line) is fitted to the calculated phase shift and shows a positive slope for $\lambda_0$=364 nm and $\lambda_0$=410 nm. **b**, Band structure showing the different harmonics. Frequency is expressed in units of $a/\lambda_0$, with $a$=152 nm the unit-cell lattice constant. m=-1,0,1 bands are shown in red, green and blue respectively. Solid dots are effective wave vectors determined from calculations as in a. The dashed horizontal line at 0.42 $a/\lambda_0$ corresponds to a free space wavelength of $\lambda_0$=364 nm.

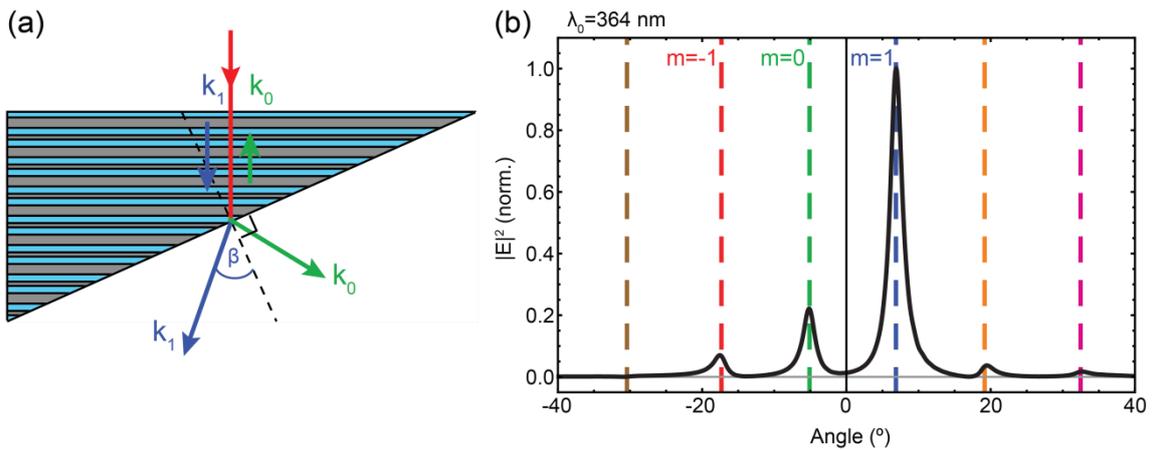

**Figure 6 a**, Sketch of metamaterial prism. The simulated prism consisted of 15 unit cells and had an output facet at 5 degrees from the interfaces. Light is normally incident on the flat side, and couples to different harmonics in the multilayer structure. Here, the wave vectors of the *m=0* and *m=1* harmonics are shown schematically. The harmonics refract with different angles due to the orientation and magnitude of the wave vector. **b**, Finite-difference time-domain simulation of the angular distribution of light refracted from a Ag/TiO$_2$ multilayer metamaterial prism at a free-space wavelength of 364 nm (layer thicknesses in Table 1). Light refracts under well-defined angles from the prism, directly probing the phase velocity inside the prism. The dominant peak, corresponding to the *m=1* harmonic, has a positive refraction angle, corresponding to a positive harmonic phase index.

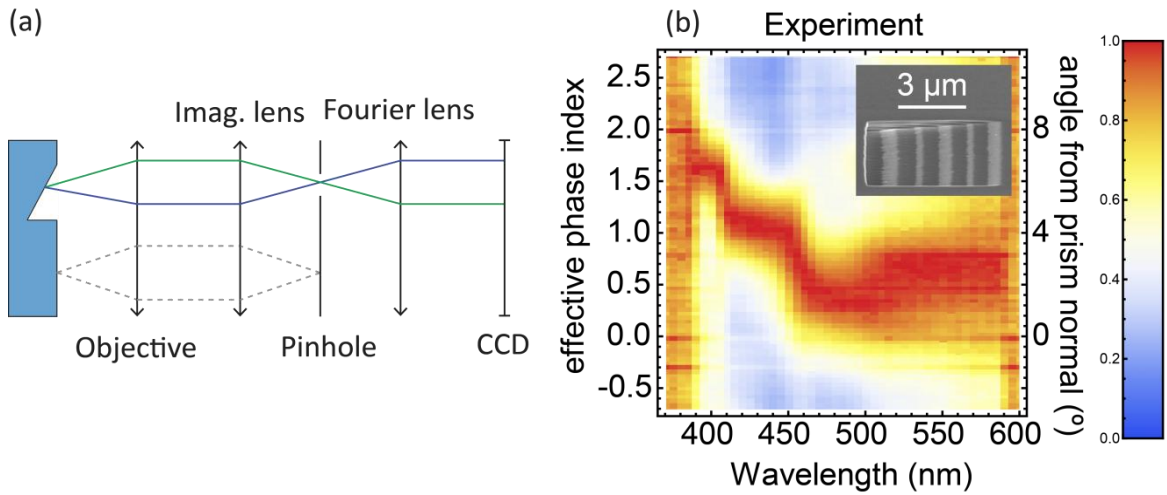

**Figure 7 a**, Sketch of the Fourier microscope used in the experiment. Light transmitted from the sample is filtered using a pinhole after an imaging lens. The back focal plane of the objective is observed by the 2D imaging CCD camera. **b**, Measured normalized light intensity as a function of refraction angle relative to the normal of the sloped side of the prism and as a function of free-space wavelength. The angle is converted to an effective phase index using Snell's law. The inset shows a SEM image of the fabricated microprism. Light refraction is governed by a positive effective index, which increases in a stepwise manner for decreasing wavelength, in agreement with the trend in Fig. 5b.